\documentclass[aps,prb,twocolumn,amssymb,showpacs]{revtex4}
\usepackage{bm}
\usepackage{graphicx}

\begin{document}

\title{Comment on ''Magnetic quantum oscillations of the conductivity 
in layered conductors''}
\author{T. Champel}
\affiliation{
Institut f\"{u}r Theoretische Festk\"{o}rperphysik,
  Universit\"{a}t Karlsruhe,
  76128 Karlsruhe, Germany }

\author{V.P. Mineev}
\affiliation{Commissariat \`{a} l'{E}nergie Atomique, 
DSM/DRFMC/SPSMS, 17 rue des Martyrs, 38054 Grenoble Cedex 9, France}

\date{\today}

\begin{abstract}
We discuss the recent theory of  Gvozdikov [Phys. Rev. B \textbf{70}, 
085113 (2004)] which aims at 
 explaining the Shubnikov-de Haas oscillations of the longitudinal 
resistivity $\rho_{zz}$ observed in the quasi-two-dimensional organic 
compound 
$\beta''-(\mathrm{BEDT-TTF})_{2}\mathrm{SF}_{5}\mathrm{CH}_{2}\mathrm{CF}_{2}\mathrm{SO}_{3}$. 
We point out that the self-consistent equations of the theory yielding
the longitudinal resistivity and the magnetic field dependence of the 
chemical
potential have been incorrectly solved.  We show that the
consideration of the self-consistent Born approximation (which
determines the relaxation rate in Gvozdikov's paper) leads in fact to
the complete absence of the longitudinal conductivity $\sigma_{zz}$ at
leading order in high magnetic fields.

\end{abstract}

\pacs{72.15.Gd, 75.20.En, 73.43.Cd, 71.18.+y}

\maketitle

Gvozdikov \cite{Gvo2004} recently derived a theory of the 
Shub\-ni\-kov-de Haas oscillations of the longitudinal conductivity 
$\sigma_{zz}$ in quasi-two-dimensional  (2D) metals in high 
perpendicular magnetic fields. He found that $\sigma_{zz}$ minima 
display a thermally activated behavior and that $\sigma_{zz}$ peaks 
are split in presence of small oscillations of the chemical 
potential $\mu$.
He also claimed that his theory can explain the giant oscillations of 
the
magnetoresistance $\rho_{zz}=\sigma_{zz}^{-1}$ observed \cite{Nam2001}
in the layered organic compound
$\beta''-(\mathrm{BEDT-TTF})_{2}\mathrm{SF}_{5}\mathrm{CH}_{2}\mathrm{CF}_{2}\mathrm{SO}_{3}$.

In this Comment, we  point out that the theoretical results have 
been incorrectly derived and that the theoretical model adopted by 
Gvozdikov 
 is inconsistent with existing experiments. 
 In the following, we first address the framework of the theory and 
the problems with the calculation of $\sigma_{zz}$. Then, we shall 
focus on the thermodynamic quantities such as the density of states 
and the chemical potential.
Finally, we shall discuss possible theoretical explanations of 
experimental observations.

\section{The longitudinal conductivity $\sigma_{zz}$}

\subsection{Model and framework}

The system under consideration is a quasi-two-dimensional metal with 
an interlayer hopping integral $t$.
In the paper, \cite{Gvo2004}  the interlayer dispersion is modelled by 
an unknown density of states with a characteristic width of the order 
of $t$.
Gvozdikov's paper \cite{Gvo2004} was aimed at deriving the 
oscillations of $\sigma_{zz}$ as a function of magnetic field
perpendicular to the layers under the regime of a weakly incoherent 
interlayer
transport.

The incoherent transport was  discussed in literature (see, e.g., Ref. 
\onlinecite{Moses}) in the cases of the so-called weakly and strongly
incoherent regimes.  The former occurs when the intralayer momentum
is conserved in the process of transfer of the electron from one layer
to another.  The incoherence is relevant when a large number of
in-plane scatterings takes place before an interlayer tunneling
occurs, i.e. when $t \ll \hbar/\tau$, where $\tau$ is the mean free time. 
Interlayer transport is then incoherent because the successive
tunneling events get uncorrelated.  This situation was first discussed
in the paper \cite{Kumar} in relationship with the conductivity in
cuprates and it gives a  Drude-type formula for the interlayer
conductivity where
the scattering time is the time of inelastic intralayer scattering.  
The strongly incoherent transport occurs when the
intralayer momentum is not conserved by tunneling and there is no
interference between the wavefunctions on adjacent layers.

In Gvozdikov's theory\cite{Gvo2004}  intralayer inelastic processes 
are apparently \cite{note2} not considered since the width
$\Gamma=\hbar/2 \tau$ of the spectral function in the layer is
determined by the elastic scattering on impurities (note that in our
definition there is a difference of a factor 2 with Gvozdikov's
notation).  The additional complicacy considered in Ref. 
\onlinecite{Gvo2004} consists of the inelastic processes at tunneling. 
The Landau levels and (implicitly) the centers of the Landau orbits
are assumed to be conserved.  This is valid provided that the
characteristic value of the energy change at tunneling, $\gamma$, is
much less than the intralayer spectral width $\Gamma$.  Under this
regime of so-called weak incoherence by Gvozdivov (that does not
correspond to the definition of weak incoherence given above), the
inelastic processes at tunneling do not play a role for the interlayer
transport and we return to a theory where only elastic scattering on
impurities is relevant.  It should be noted that there is no clear
distinction between $t$ and $\gamma$ in Ref. 
\onlinecite{Gvo2004}.

The derivation of the longitudinal conductivity $\sigma_{zz}$ for 
elastic impurity scattering  and a cosinelike interlayer energy 
dispersion was already 
reported in the Refs \onlinecite{Gri2003,Cha2002,Cha2004}.  The paper 
\cite{Gri2003} was 
interested in the regime $\hbar \Omega \leq t$ (here $\Omega$ 
is the cyclotron frequency), while the papers \cite{Cha2002,Cha2004} 
were
concerned with the 2D limit $\hbar \Omega \gg t$ in connection 
with the experiment. \cite{Nam2001}

\subsection{Derivation of $\sigma_{zz}$}
The longitudinal conductivity  
$\sigma_{zz}$ is derived within the Kubo formula written in the 
formalism of the Green functions. Vertex corrections have been 
neglected in Ref. \onlinecite{Gvo2004} which means that a point-like 
impurity model has been implicitly assumed.
Within this model the self-energy part entering into 
the expression of the disorder averaged Green functions is 
independent of the 
Landau levels index $n$. This fact allows to compute analytically the 
sum over $n$ using the Poisson summation formula. 

Gvozdikov has derived an expression for $\sigma_{zz}$,  Eqs. 
(21)-(22).  In addition to the condition $t \ll \Gamma$, the 2D limit 
$\hbar \Omega \gg t$ is required for its validity.
Under the same inequalities, one can rewrite these equations 
(21)-(22) as

\begin{equation}
\sigma_{zz}=\int dE \left(-f'(E)\right) \left[ 
\sigma_{B}+\sigma_{Q}\right] \label{Eq1},
\end{equation}
where
\begin{eqnarray}
\sigma_{B}&=&\sigma_{\tau} \frac{\lambda_{0}}{\lambda(E)} 
S(\lambda(E),E),  \label{Eq2} \\
\sigma_{Q}&=& -\sigma_{\tau} \lambda_{0} \frac{\partial 
S(\lambda(E), E)}{\partial \lambda}. \label{Eq3}
\end{eqnarray}
Here $f(E)$ is the Fermi-Dirac distribution function which depends on 
the chemical potential $\mu$, 
$\sigma_{\tau}$ is the conductivity at zero magnetic field, 
$\lambda(E)=\pi/\Omega \tau(E)$, 
$\lambda_{0}= \pi/\Omega \tau_{0}$ [$\tau(E)$ and $\tau_{0}$ are 
respectively the {\em elastic} quasi-particle lifetimes at energy 
$E$ in the presence and absence of magnetic field] and the 
function $S$ is defined by
\begin{equation}
S(\lambda(E), E)=\frac{\sinh 
\lambda(E)}{\cosh \lambda(E)+\cos(2 \pi 
E/\hbar \Omega)}. \label{Eq4}
\end{equation}

Formulas (1)-(4) are incomplete without a prescription for 
calculating the function $\lambda(E)$.
To proceed further, Gvozdikov applied the self-consistent Born
approximation (SCBA) to the quasi-2D spectrum.  This approximation
stipulates the proportionality between the density of states
$N(E)$ and $\tau^{-1}(E)$, i.e.,
\begin{equation}
\frac{\tau_{0}}{\tau(E)}=\frac{N(E)}{N_{0}}.
\end{equation}
Here $N_{0}$ is the density of states of the quasi-2D system at zero 
magnetic field. The density of states $N(E)$ under magnetic 
field is expressed as a Fourier series and is a function of the 
lifetime $\tau(E)$. In the 2D limit ($\hbar \Omega \gg t$) and for $t 
\ll \Gamma$, it is 
possible to sum 
up this series and the relation (5) leads to the following 
self-consistent equation for $\lambda(E)$:
\begin{equation}
\lambda(E)=\lambda_{0} \, S\left(\lambda(E), 
E \right) \label{Eq6}.
\end{equation}
Equation (\ref{Eq6}) yields straightforwardly the fact that the term 
$\sigma_{B}=\sigma_{\tau}$ [see Eq. (\ref{Eq2})], i.e., this latter 
does not oscillate with the magnetic field. Gvozdikov got then that  
the oscillations in the strong 2D regime arise from the term  
$\sigma_{Q}$ only.
Here we want to stress that 
this result \cite{Gvo2004} has been incorrectly derived from the 
self-consistent Eq. (\ref{Eq6}).
All the subsequent results based on the formula (28) for 
$\sigma_{zz}$   in  Ref. 
\onlinecite{Gvo2004} 
are accordingly incorrect.
Indeed, differentiating the formula (\ref{Eq6}) with respect to 
$\lambda$, we obtain that $\lambda_{0} \partial S/\partial \lambda 
=1$. Substituting this result into Eq. (\ref{Eq3}) we 
straightforwardly have $\sigma_{Q}=-\sigma_{\tau}$, which means that 
in frame of SCBA the total conductivity
\begin{equation}
\sigma_{zz}=0
\label{Eq45}
\end{equation}
in the limit $\hbar \Omega \gg t$.
A non-zero expression for $\sigma_{zz}$ may be obtained by
considering next order corrections in the parameters $t/\hbar \Omega$
and $t/\Gamma$.

It is worth noting that the Eqs. (\ref{Eq1})-(\ref{Eq4}) are 
exactly the 
same as Eqs. (19)-(21) given in the work \cite{Cha2002} 
considering a cosinelike dispersion relation with respect to the 
interlayer momentum. 
The reason for this is that  the specific form of the interlayer 
density of states is unimportant under the conditions $t \ll \hbar 
\Omega$ and $t \ll \hbar /\tau$. The important difference between the 
two theories is that in Ref.  \onlinecite{Cha2002}
the total (not just the intralayer) spectral width due to elastic
scattering on impurities $\Gamma=\hbar/2\tau$ is postulated as being 
constant, i.e. energy independent: the SCBA leading to Eq.  
(\ref{Eq6}) and Eq. (\ref{Eq45}) is not considered as an inherent
property of the transport theory. 

 We would like to stress that Eq. (\ref{Eq45}) is a consequence of 
the application of SCBA in high magnetic fields only. In smaller 
magnetic fields $\hbar \Omega \leq t$, the same SCBA yields 
\cite{Gri2003} small oscillations of $\sigma_{zz}$ around the zero 
field value which is nonzero due to 3D elastic scattering on 
impurities.

\section{Thermodynamic quantities} 

\subsection{Density of states}

It is important to note that Eq. (\ref{Eq6})  determines as well the 
function  $\lambda(E)$ (or the lifetime 
$\tau(E)=\pi/\Omega \lambda(E)$)  as the 
density of states $N(E)=N_{0} 
\lambda(E)/\lambda_{0}$.
It has been recognized in the Section IV of Ref. \onlinecite{Cha2002} 
that the self-consistent 
equation (\ref{Eq6}) for the density of states is exactly the same 
equation 
as one would obtain using the SCBA with a strict 2D spectrum, which 
has has been well known for a long time. \cite{Ando74,Ando82}
The approximation
consisting in replacing $\lambda(E)$
everywhere by the peak value $\lambda(E_{n})\approx \sqrt{2
\lambda_{0}}$ as done in Ref.  \onlinecite{Gvo2004} is inconsistent 
with 
Eq. (\ref{Eq6}).  In
fact, the self-consistent Eq.  (\ref{Eq6}) yields a function
$\lambda(E)$ which strongly oscillates with $E$, especially in high
magnetic fields.  When $\lambda_{0} \leq 2$, $\lambda(E)$ even
vanishes between Landau levels energy positions $E_{n}=(n+1/2)\hbar
\Omega$, which means that the density of states splits into separate
bands centered on the energies $E_{n}$ (see Refs. 
\onlinecite{Ando74,Ando82}).  Going beyond SCBA, it is possible to
calculate more accurately the tails of the bands to avoid the
unphysical cutoff produced by Eq.  (\ref{Eq6}).  Moreover, next
corrections due to the interlayer hopping, of the order of $t/\hbar
\Omega$ and $t/\Gamma$, may yield finite but small values for $N(E)$
between Landau levels.  However, in any cases, the model adopted by
Gvozdikov automatically implies sharp peaks for $N(E)$ when $E \approx
E_{n}$ and the quasiabsence of states between Landau levels for
$\Omega \tau_{0} > \pi/2$.

\subsection{Chemical potential}

Eq. (23) of Ref. \onlinecite{Gvo2004} [reproduced here with Eq. 
(\ref{Eq6})] completely  determines the density of states and thus 
allows to compute the chemical potential $\mu$ as a function of 
magnetic field. Therefore, there is no freedom for the choice of some 
phenomenological model describing the chemical potential oscillations 
in accordance with the experiments, as surprisingly done in the 
Section 4 of Ref. \onlinecite{Gvo2004}.

 If the density of states consists of sharp bands centered on the 
Landau levels, one expects that $\mu$ is pinned to a value close to 
$(n+1/2) \hbar \Omega$ for most of the range in magnetic fields and 
drops suddenly to another Landau band once a Landau band becomes 
filled or empty. This process gives rise to strong magnetic quantum 
oscillations of the chemical potential. This expectation is 
physically inconsistent with the experimental observation of an 
inverse sawtooth profile of the magnetization oscillations in the 
compound 
$\beta''-(\mathrm{BEDT-TTF})_{2}\mathrm{SF}_{5}\mathrm{CH}_{2}\mathrm{CF}_{2}\mathrm{SO}_{3}$ 
,
\cite{Wos2000} that implies \cite{Cha2001b} the presence of 
negligibly small oscillations of $\mu$. Furthermore, the experiments 
on thermodynamic and transport quantities\cite{Wos2000,Nam2001} 
rather suggest a significant amount of states between the Landau 
levels for $\Omega \tau_{0} \sim 1$ in contradiction with the result 
obtained within the self-consistent Born approximation.

Now, we discuss the different forms possible for the chemical 
potential oscillations under the assumption of a constant lifetime 
(as in Section 4 of Ref. \onlinecite{Gvo2004}), because they have 
been incorrectly presented and used in Ref. 
\onlinecite{Gvo2004}.  For an energy independent lifetime 
$\bar{\tau}$, the equation obeyed at
zero temperature by the chemical potential $\mu$ in the 2D limit  can 
be explicitly calculated \cite{Cha2001a,Cha2001b}
\begin{equation}
\mu=E_{F}+\frac{\hbar \Omega}{\pi} 
\arctan\left(\frac{\sin(2 \pi \mu/\hbar \Omega)}{e^{\nu}+\cos(2 
\pi \mu/\hbar \Omega)}
\right) \label{chem}
\end{equation}
where $\nu=\pi/\Omega \bar{\tau}$ and $E_{F}$ is the 
Fermi energy at zero magnetic field. 
As pointed out by Grigoriev, \cite{Gri2001} it is possible to invert 
this self-consistent formula (\ref{chem}) to obtain
\begin{equation}
\mu=E_{F}+\frac{\hbar \Omega}{\pi} 
\arctan\left(\frac{\sin(2 \pi E_{F}/\hbar 
\Omega)}{e^{\nu}-\cos(2 \pi E_{F}/\hbar \Omega)} 
\label{correct}
\right).
\end{equation}
When $\nu \ll 1$, the density of states within this model consists of 
sharp peaks centered around the Landau levels (as in the SCBA),
and 
this Eq. (\ref{correct}) yields a staircase dependence for $\mu/\hbar 
\Omega$ as a function of magnetic field. Correspondingly, the 
oscillating 
part $\tilde{\mu}/\hbar \Omega=(\mu-E_{F})/\hbar 
\Omega$ exhibits a sawtooth dependence (with the so-called direct 
shape). \cite{Cha2001b}

 Within the phenomenological model assuming a reservoir of states and 
a constant lifetime, the equation for $\mu$ now takes the form 
\cite{Cha2001b}

\begin{equation}
\mu=E_{F}+\frac{\hbar \Omega}{\pi(1+R)} 
\arctan\left(\frac{\sin(2 \pi \mu/\hbar \Omega)}{e^{\nu}+\cos(2 
\pi \mu/\hbar \Omega)}
\right) \label{res}
\end{equation}
where $R$ is a dimensionless parameter measuring the strength of the 
reservoir. Equation (\ref{res}) embraces naturally the former model 
with $R=0$.
For an important reservoir, $R \gg 1$, the oscillating part 
$\tilde{\mu}$ is extremely small ($\tilde{\mu} \ll \hbar 
\Omega$), implying that $\mu$ is almost fixed to the zero field 
value $E_{F}$. In this particular case, it is then 
justified to replace $\mu$ by $E_{F}$ in the right-hand 
side of this Eq. (\ref{res}) to obtain
\begin{equation}
\mu \approx E_{F}+\frac{\hbar \Omega}{\pi R} 
\arctan\left(\frac{\sin(2 \pi E_{F}/\hbar 
\Omega)}{e^{\nu}+\cos(2 \pi E_{F}/\hbar \Omega)}
\right) .
\end{equation}
Then, $\tilde{\mu}/\hbar \Omega$ exhibits the (so-called) inverse 
sawtooth waveform when $\nu \ll 1$, but with a negligibly small 
amplitude (because $R \gg 1$). On the contrary, the magnetization 
oscillations which are proportional to  $R \tilde{\mu}/\hbar 
\Omega$ are not limited in amplitude by the reservoir parameter 
$R$ (see Ref. \onlinecite{Cha2001b}).

In his discussion of the  effects of the chemical potential 
oscillations on $\sigma_{zz}$, Gvozdikov \cite{Gvo2004} has used Eq. 
(\ref{chem}) (i.e. no reservoir) with $\mu$ replaced by 
$E_{F}$ in the right-hand side of the equation.
We want to stress that it is generally mathematically incorrect to 
replace $\mu$ by
$E_{F}$ \cite{Cha2001a,Cha2001b} in the right-hand side of
Eq.  (\ref{chem}) especially when  $\nu \ll 1$ [e.g., compare the 
resulting equation and Eq. 
(\ref{correct})].
An additional remark is that Eq. (31) for $\mu$ considered in
Ref.  \onlinecite{Gvo2004} [Eq.  (\ref{chem}) with a minus
instead of a plus between the two terms in the right-hand side] to
describe the direct sawtooth case has no  basis at all.

\section{Theoretical approaches}

As already mentioned above, the theory in Ref. \onlinecite{Cha2002}, 
also aimed at explaining the oscillations of $\sigma_{zz}$ reported 
in Ref. \onlinecite{Nam2001}, is based on the same Eqs. (1)-(4) for 
$\sigma_{zz}$. The difference with Gvozdikov's theory rests on the 
function $\lambda(E)$ [or equivalently on the lifetime $\tau(E)$], 
which is (phenomenologically) assumed to be constant. 
Within this model  both contributions 
$\sigma_{B}$ and $\sigma_{Q}$ oscillate with magnetic fields. The 
thermal activation of $\sigma_{zz}$ minima stems from a cancellation 
at zero order in $t/\hbar \Omega$ of these two contributions 
between the Landau levels. \cite{Cha2002,Cha2004} Contrary to SCBA, 
the cancellation does not occur at the Landau levels.

Now, we would like to address one alternative explanation to the 
experiment. \cite{Nam2001}
Owing to some similitudes, it is tempting to bring together the 
present problem in a layered conductor and the problem encountered in 
the theory of the quantum Hall effect  \cite{Hajdu}. 
At the experimental side, in this quasi-2D conductor and in the 2D 
electron gases under the regime of the integer quantum Hall effect, a 
thermal activation is observed: in the first system it concerns the 
minima of 
the interlayer conductivity $\sigma_{zz}$, \cite{Nam2001} and in the 
second system 
the minima of the longitudinal conductivity $\sigma_{xx}$. 
\cite{IQHE} 
The puzzling point is that the density of states extracted from 
different measurements of the thermodynamic quantities points out a 
rather large amount of states between the Landau peaks in the layered 
organic conductor \cite{Wos2000,Nam2001} and in the 2D electron gases 
under the regime of the integer quantum Hall effect  \cite{Gor,Eis} 
which is typically observed when $\Omega \tau_{0} \geq 1$. 
 As a result,
we have to face the same difficulty, i.e., to capture transport and 
thermodynamic properties within a complete and consistent microscopic 
theory.
In the regime of the integer quantum Hall effect,
it is believed that most of the states of the 2D layer are localized 
by the disorder and act as a reservoir which almost fixes  the 
chemical potential. \cite{IQHE} The core of the extended states is 
located at the energies $E_{n}$. At finite but not too low 
temperatures,  the main means of conduction when the chemical 
potential sits between the Landau levels is the thermal excitation of 
quasiparticles at the edges of the mobility gap. \cite{IQHE}
We could presume the presence of similar localized states in the 
layers responsible 
for the thermal activation of $\sigma_{zz}$ minima at high magnetic 
fields  in the compound 
$\beta''-\mathrm{(BEDT-TTF)}_{2}\mathrm{SF}_{5}\mathrm{CH}_{2}\mathrm{CF}_{2}\mathrm{SO}_{3}$.

For the moment there exists no proper quantum calculations for 
$\sigma_{xx}$ in the quantum Hall effect regime, \cite{Hajdu} so that 
this scenario for $\sigma_{zz}$ only rests on pure qualitative
considerations.
In fact, the principal theoretical difficulty is  the microscopic 
treatment of the impurity scattering effects on the Landau levels 
quantization.
For this purpose,
it is worth noting that
the consideration of a model of impurity potential with zero-range is 
unphysical in high magnetic fields B since in the limit $B \to 
\infty$ the magnetic length $l_{B}=(\hbar c/eB)^{1/2}$ always becomes 
smaller than the correlation radius of the potential. 
It has been shown  \cite{Rai1993} for the 2D systems that the SCBA is 
no more valid when the range of the potential exceeds $l_{B}$  so 
that the investigation of impurity effects within the usual 
perturbation theory becomes very complicated. However, it seems that 
even this ingredient which consists to consider a finite range for 
the impurity potential is not sufficient to explain the presence of a 
background reservoir of states. \cite{Spi1997} Apparently, new ideas 
are needed.

\section{Conclusion}

The results reported in the paper \cite{Gvo2004}  are based on the
inconsequent use of SCBA. The proper SCBA application at leading
order in high magnetic field leads to the zero conductivity
$\sigma_{zz}¥=0$ and to strong oscillations of the density of states 
and of the 
chemical potential.  Both latter effects can be suppressed in the
presence of a significant number of states between Landau levels.  An 
explanation for the thermally activated behavior
of $\sigma_{zz}$ minima could be the presence of localized states as  
in  the integer quantum Hall effect.  However, a concrete
self-consistent magnetotransport theory going beyond SCBA and including
these localized states is still missing.

\section*{Acknowledgments}
T.C. acknowledges financial support from the Deutsche 
Forschungsgemeinschaft within the Center for Functional 
Nanostructures.

\end{document}